\pdfoutput=1

\documentclass[11pt]{article}

\usepackage{acl}

\usepackage{times}
\usepackage{latexsym}

\usepackage[T1]{fontenc}

\usepackage[utf8]{inputenc}

\usepackage{microtype}

\usepackage{inconsolata}

\usepackage{amsmath,amsthm,amssymb,amsfonts}
\usepackage{booktabs}
\usepackage{multirow}
\usepackage{enumerate}
\usepackage{enumitem}
\usepackage{epigraph}

\usepackage{pgfplots}
\pgfplotsset{compat=1.18}
\usepackage{pgfplotstable}
\usepackage{makecell}
\usepackage{color}
\usepackage{colortbl}  
\usepackage{xcolor}
\usepackage{array} 
\usepackage{caption}
\usepackage{graphicx}
\usepackage{float} 
\usepackage{subfigure}
\usepackage{subcaption}
\usepackage{parskip}

\usepackage[capitalise]{cleveref}
\crefname{section}{§}{§§}
\title{Unsupervised Query Routing for Retrieval Augmented Generation}

\author{
 \textbf{Feiteng Mu\textsuperscript{1}\thanks{Work done during internship at Alibaba Inc.}},
 \textbf{Liwen Zhang\textsuperscript{2}},
 \textbf{Yong Jiang\textsuperscript{2}\thanks{Corresponding author.}},\\
 \textbf{Zhen Zhang\textsuperscript{3}},
 \textbf{Wenjie Li\textsuperscript{1}},
 \textbf{Pengjun Xie\textsuperscript{2}},
 \textbf{Fei Huang\textsuperscript{2}},
\\
 \textsuperscript{1}The Department of Computing, The Hong Kong Polytechnic University, Hong Kong\\
 \textsuperscript{2}Institute for Intelligent Computing, Alibaba Group\\
 \textsuperscript{3}College of Software, Nankai University\\
\texttt{\{csfmu,cswjli\}@comp.polyu.edu.hk,\{yongjiang.jy,zlw439616\}@alibaba-inc.com}
}

\begin{document}
\maketitle

\begin{abstract}
Query routing for retrieval-augmented generation aims to assign an input query to the most suitable search engine. Existing works rely heavily on supervised datasets that require extensive manual annotation, resulting in high costs and limited scalability, as well as poor generalization to out-of-distribution scenarios. To address these challenges, we introduce a novel unsupervised method that constructs the "upper-bound" response to evaluate the quality of retrieval-augmented responses. This evaluation enables the decision of the most suitable search engine for a given query. By eliminating manual annotations, our approach can automatically process large-scale real user queries and create training data. We conduct extensive experiments across five datasets, demonstrating that our method significantly enhances scalability and generalization capabilities.
\end{abstract}

\section{Introduction}

Retrieval Augmented Generation (RAG) \cite{lewis2020retrieval,gao2023retrieval} 
typically begins with a retrieval phase where user queries are scoured through search engines like Google and Bing to gather relevant background documents before engaging large language models (LLMs) \cite{gpt4,qwen2}. In today's digital landscape, even among major comprehensive search engines, each has its strengths and weaknesses, offering unique features and capabilities. Consequently, \textit{Query Routing} \cite{sugiura2000query,wang2024resllm,mu2024adaptive} seeks to direct queries to the most suitable search engines to optimize performance and reduce costs.

Current query routing methods \cite{shnitzer2023large,mu2024adaptive} primarily use open-sourced \textit{(query, answer)} paired data to create training labels. 
Specifically, by leveraging these gold-standard answers, they assess the quality of retrieval-augmented responses for queries routed through different search engines, thereby identifying the most effective engine for each query. 
However, the reliance on annotated paired data presents challenges such as limited diversity and high annotation costs, which significantly restrict scalability. Moreover, there exists a pronounced discrepancy between the distribution of \textit{public datasets} and \textit{real user queries}, which severely impacts the out-of-distribution generalization \cite{shnitzer2023large} of these methods, limiting their effectiveness in real-world applications. To enable models to generalize effectively to actual user queries, it is essential to train them directly on real user queries.
Nevertheless, real user queries often lack gold-standard answers, presenting a major challenge in determining the appropriate search engine assignment, particularly when \textit{no definitive answer is available}.

\begin{table}
\centering\small
\begin{tabular}{@{}lll@{}}
\toprule
Search Strategies & Qwen2-max & GPT4 \\ \midrule
No-RAG & 2.189 & 2.014 \\ \midrule
\rowcolor{gray!10}
\multicolumn{3}{c}{\textit{Using one search engine}} \\
Quark & 2.651 & 2.456 \\
Bing & 2.615 & 2.390 \\
Google & 2.610 & 2.420 \\ \midrule
\rowcolor{gray!10}
\multicolumn{3}{c}{\textit{Using two search engines}} \\
Quark+Bing & 2.689 & 2.495 \\
Quark+Google & 2.701 & 2.486 \\
Bing+Google & 2.674 & 2.478 \\ \midrule
\rowcolor{gray!10}
\multicolumn{3}{c}{\textit{Using three search engines}} \\
Quark+Bing+Google & \textbf{2.727} & \textbf{2.521} \\ \bottomrule
\end{tabular}
\caption{RAG results across various search strategies. We merge all test sets and calculate average end2end QA scores, i.e., Correctness, on the merged examples. No-RAG denotes that LLMs respond to the query without retrieval. Scores with \textbf{bold} denote the best results.}
\label{tab:intro_searchmethod}
\end{table}

To address this challenge, we must transform our problem into a labelling-free form.
Since query routing fundamentally involves identifying the optimal assignment within the limited computing resources, such as inference time and budget, we can envision an ideal situation in which we possess infinite computing resources.
In this case, an intuitive and obvious strategy is to search a query across all search engines, then merge multi-sourced documents for RAG\footnote{For simplicity, we denote the generated response when retrieving from multiple tools as the multi-sourced response. Correspondingly, we denote the generated response when retrieving from only one tool as the single-sourced response.}.
Due to the complementarity of different search engines \cite{mu2024adaptive}, the merged multi-sourced documents should have a better recall than the single-sourced documents. 
That is, multi-sourced responses are likely to exhibit higher quality than single-source responses.  
To support this assertion, we calculate the RAG results across various search strategies and different LLMs, as illustrated in Table \ref{tab:intro_searchmethod}. We find that utilizing more search engines consistently leads to improved results. This trend is observed across multiple LLMs.
Based on this prior knowledge, we can view the multi-sourced responses as an upper bound for unsupervised query routing. 
Now, let us return to the practical situation of limited computing resources.
To evaluate the quality of a single-sourced response, we can compare it to the corresponding multi-sourced response, which serves as an upper bound. 
By assessing the similarity between the single-sourced response and its upper bound, we are able to determine its quality. If the single-sourced response shows a high level of similarity to its upper bound, this suggests that the query should be assigned to the corresponding search engine.
This chain of thought motivates us to develop an unsupervised query routing method that directly leverages real user data to generate training labels in an automated manner, eliminating the need for manual annotation.

In this work, we propose a novel unsupervised query routing method that directly creates training data without human annotations.
Our method mainly consists of four steps. 
In the first step, we search every user query in only a single search engine to build single-sourced responses.
In the second step, we search every user query across all available search engines to establish multi-sourced responses, which are used as the upper bound for unsupervised query routing. 
Next, given a query and its corresponding single-sourced responses, we use several automated metrics to evaluate the quality of the single-sourced responses, allowing us to determine which search engine best fits the query. Finally, by constructing training data automatically, we train our routing model. 
Our method operates without the need for manual annotation, enabling the automation of large-scale processing of real-world user queries and facilitating the construction of training data. Consequently, our approach can be effectively scaled and has a strong generalization capability for real user queries.

In summary, our contributions are as follows.
(1) We introduce a novel unsupervised query routing method capable of automatically processing large-scale, unlabeled real user data for model training. This simple yet effective approach significantly reduces annotation costs and offers excellent scalability potential.
(2) As far as we know, we are the first to explore unsupervised query routing within the RAG scenario. Our method serves as a universal framework, holding significant promise for providing valuable insights into the field of tool learning.
(3) We validate our method across five datasets in a non-i.i.d. setting. The experiments demonstrate that our approach exhibits \textbf{i)} outstanding generalization ability, \textbf{ii)} excellent scalability, and \textbf{iii)} consistent effectiveness across multiple LLMs.

\section{Related Work}

\paragraph{Query Routing for LLMs}
Nearly monthly released LLMs \cite{qwen2,dubey2024llama} are trained with various data, resulting in a variety of strengths and weaknesses in versatile downstream tasks \cite{jiang2023llm}. 
Therefore, researchers \cite{narayanan2023tryage,lu2023routing,liu2024optllm,shnitzer2023large,chen2023frugalgpt,vsakota2024fly} are devoted to learning the optimal assignment of queries to different LLMs.
The basic idea is to leverage the complementarity of various LLMs to achieve superior performance compared to any single model across a range of tasks.
Notably, these methods typically require substantial amounts of annotated training data, leading to high costs for label collection. In contrast to these approaches, we explore query routing across different search engines and propose an unsupervised method, thereby eliminating the need for manual annotation and significantly reducing annotation costs.

\paragraph{Query Routing for Search Engines}
Early research focused on topic-specific search engines \cite{manber1997search,sugiura2000query}, which often suffer from limited coverage. Alternatively, other conventional systems \cite{gravano1999gloss,liu1999query} target on general-purpose search engines, but they require access to the complete internal databases associated with each engine.
To address this issue, \cite{mu2024adaptive} recently utilized annotated (query, answer) data to train neural models that can find the most appropriate general-purpose search engine for a given query. However, these supervised methods rely heavily on annotated data.
In contrast, our method does not require manual annotation and directly uses real user queries for training, thus offering promising scalability and generalization capabilities.


\begin{figure*}[!htb]
    \centering
    \includegraphics[width=0.95\linewidth]{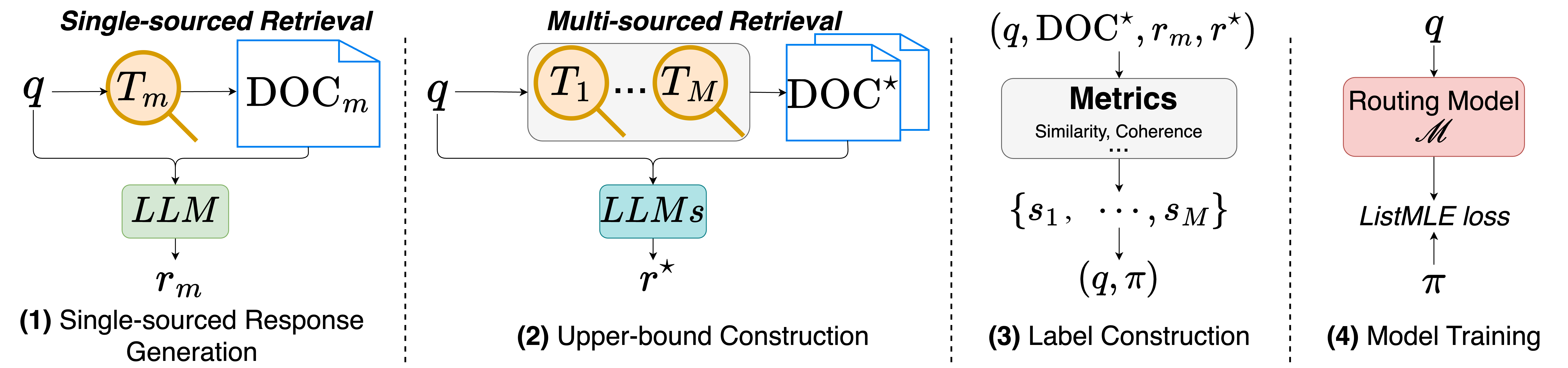}
    \caption{The overall framework of our method, which mainly consists of four steps. We automatically construct supervision information for training our routing model.}
    \label{fig:framework}
\end{figure*}

\section{Preliminaries}
Given a user query $q$ and $M$ search tools $ \{T_m\}_{m=1}^M $, the core task of query routing is to create annotated training data in the form $ (q, \pi) $, where $ \pi $ represents the search engine index indicating which search engine should be selected to address the query.

\subsection{Briefing of Previous Supervised Methods}
\label{sec:super_works}

Existing supervised methods \cite{shnitzer2023large,mu2024adaptive} rely heavily on annotated (query, answer) paired data. Specifically, using each search tool $T_m$, they are able to obtain the retrieval-augmented response $r_m$, respectively. By calculating the similarity between $r_m$ and the gold answer $r_{\text{gold}}$, they can obtain the labeled index: 
\begin{equation}
\setlength{\belowdisplayskip}{4pt} 
\setlength{\abovedisplayskip}{4pt}
    \pi=\arg\max_m \text{similarity}(r_m, r_{\text{gold}}).
    \label{eq:super_label}
\end{equation}
However, as mentioned earlier, public datasets tend to be relatively small in scale, lack diversity, and significantly differ from actual user queries, which restricts their scalability and generalization ability.

\subsection{Problem Formulation} 

We devise an unsupervised method that relies solely on user queries and search tools to generate training data.
Formally, we first obtain the single-sourced responses using every search tool $T_m$:
\begin{equation}
\setlength{\belowdisplayskip}{4pt} 
\setlength{\abovedisplayskip}{4pt}
    \begin{split}
        \textbf{DOC}_{m} &= T_m(q)\\
    r_m &= \text{LLM}(q,\textbf{DOC}_{m}),
    \label{eq:single_rag}
    \end{split}
\end{equation}
where $\textbf{DOC}_{m}$ denotes the single-sourced documents retrieved from $T_m$.
Then, by aggregating retrieved documents from multiple search tools, we obtain the upper-bound multi-sourced response:
\begin{equation}
\setlength{\belowdisplayskip}{4pt} 
\setlength{\abovedisplayskip}{4pt}
    \begin{split}
        \textbf{DOC}^{\star}& = T_1(q) || \cdots || T_M(q) \\
    r^{\star} &= \text{LLMs}(q,\textbf{DOC}^{\star}),
    \label{eq:multi_rag}
    \end{split}
\end{equation}
where $\cdot || \cdot$ denotes the merging operation, $\textbf{DOC}^{\star}$ denotes the multi-sourced documents, and $r^{\star}$ denotes the upper-bound multi-sourced response.
Finally, by using the upper-bound $r^{\star}$ as the anchor, we evaluate the quality of every single-sourced response, which allows us to obtain the training label for $q$:
\begin{equation}
\setlength{\belowdisplayskip}{4pt} 
\setlength{\abovedisplayskip}{4pt}
\pi= \text{Metrics}(q,\text{DOC}^\star,r_m, r^\star),
    \label{eq:unsuper_label}
\end{equation}
where the details of used metrics are shown in \cref{sec:label_metrics}.
By comparing Equation \ref{eq:super_label} with Equation 
\ref{eq:unsuper_label}, it can be found that through a series of transformations, we use upper-bound multi-sourced $r^\star$ instead of the manually annotated $r_{\text{gold}}$, thus achieving the goal of unsupervised query routing.

\section{Methodology}

The overall framework of our method is shown in Figure \ref{fig:framework}. 
Our method mainly consists of four steps: (1) \textbf{Single-sourced Response Generation} which generates single-sourced responses for each search tool; (2) \textbf{Upper-bound Construction} which constructs the multi-sourced responses; (3) \textbf{Label Construction} which utilizes the single-sourced response and its corresponding multi-sourced upper-bound to create training labels; and (4) \textbf{Model Training} where we train our routing model.

\subsection{Single-sourced Response Generation}
\label{sec:singlesourced_rag}
As stated in Equation \ref{eq:single_rag}, for each tool $T_m$, we search $q$ within $T_m$ to retrieve $\textbf{DOC}_{m}$ which consists of $k$ documents.  
Next, we compile the query $q$ and the retrieved document $\textbf{DOC}_{m}$ into a prompt, which is then fed into a LLM to generate the single-sourced response $r_m$.
The prompt is shown in Figure \ref{fig:single_rag_prompt}. For the LLM, we employ Qwen2-max\footnote{https://dashscope.aliyuncs.com/compatible-mode/v1} \cite{Yang2024Qwen2TR} to generate responses.
This process is repeated for each search tool  $T_m$, resulting in a collection of single-sourced responses $\{r_m\}_{m=1}^M$.

\subsection{Upper-bound Construction}
Given $q$ and $\{T_m\}_{m=1}^M$, we can obtain $\textbf{DOC}_{m}$ for each $T_m$ (Equation \ref{eq:single_rag}), respectively. The question is how to retain $k$ documents from $k*M$ documents to achieve a fair comparison to \cref{sec:singlesourced_rag}. Previous works generally utilize the text reranker \cite{li2023towards,bge_m3} to retain top-ranked documents. However, these rerankers require extra training and may not generalize well to black-box search engines. 
Therefore, we opt to forgo the re-ranking process and instead utilize the inherent sorting algorithm of the search engine directly.
Specifically, for $\textbf{DOC}_{m}$ that contains the ordered $k$ documents, we only use the top-ranked $k/m$ documents. That is, for $\{\textbf{DOC}_{m}\}_{m=1}^M$, we finally reserve the multi-sourced documents $\textbf{DOC}^\star$ which contains $k$ documents. Then, we pass $q$ and $\textbf{DOC}^\star$ to LLMs to generate the multi-sourced response $r^\star$ (Equation \ref{eq:multi_rag}).
Notably, we consider multiple LLMs, including Qwen2-max and GPT4, for generating $r^\star$ to improve the robustness of $r^\star$.
This is motivated by the previous works of text generation \cite{Bahdanau2014NeuralMT,Sutskever2014SequenceTS}  in which the gold references usually contain multiple texts.

\subsection{Label Construction}
\label{sec:label_metrics}
We devise two metrics, i.e., similarity and coherence scores, to assess the quality of $\{r_m\}_{m=1}^M$. Considering multiple scores for each $r_m$, we adopt a normalization schema to get the overall score, so that obtain the training labels.

\paragraph{Similarity Score} We first use BertScore \cite{Zhang2019BERTScoreET} to evaluate the similarity between $r_m$ and its upper-bound $r^\star$:
\begin{equation}
\setlength{\belowdisplayskip}{4pt} 
\setlength{\abovedisplayskip}{4pt}
    s_{m}^\text{Bert}=\text{BertScore}(r_m,r^\star).
\end{equation}
The intuition is that the higher the value of $s_{m}^\text{Bert}$, the better the quality of $r_m$.

\paragraph{Coherence Score} Besides Similarity, we also care about the coherence between the query and the single-sourced response, i.e., whether $r_m$ can effectively solve $q$. Therefore, we use a LLM, e.g., Qwen2-max, to determine which one better solves $q$ in a pair of single-sourced responses.
Without loss of generality, we define the paired single-sourced responses as $r_a$ and $r_b$. Taken $q$, the multi-sourced documents $\textbf{DOC}^\star$, $r_a$, and $r_b$ as inputs, the LLM determines the one with better quality\footnote{The prompt is shown in Appendix \ref{app:prompts}, Figure \ref{fig:nli_prompt}.}:
\begin{equation}
\setlength{\belowdisplayskip}{4pt} 
\setlength{\abovedisplayskip}{4pt}
    r_a \quad \text{or}\quad r_b \longleftarrow \text{LLM}(q,\textbf{DOC}^\star,r_a,r_b).
\end{equation}
Through $\frac{M*(M-1)}{2}$ comparisons, we can obtain the overall ranking relationship (in ascending order) among $\{r_m\}_{m=1}^M$\footnote{We discard the illegal ranking output by LLMs (like "$A>B,B>C,C>A$").}.
We define the coherence score $s_m^\text{Coh}$ as the index of $r_m$ in the overall ranking. The larger the index, the higher the score.

\paragraph{Overall Score}
Having obtained $s_{m}^\text{Bert}$ and $s_m^\text{Coh}$ of $r_m$, we devise a normalization schema to get the overall score of $r_m$, which helps to mitigate the impact of varying dimensions of multiple metrics. For example, we calculate the mean $\mu^\text{Bert}$ and standard deviation $\sigma^\text{Bert}$ of the similarity score across all examples.
Then, the normalized similarity score and coherence score are calculated as follows:
\begin{equation}
\small
\setlength{\belowdisplayskip}{4pt} 
\setlength{\abovedisplayskip}{4pt}
    \overline{s_{m}^\text{Bert}}=\frac{s_{m}^\text{Bert}-\mu^\text{Bert}}{\sigma^\text{Bert}},\quad
    \overline{s_{m}^\text{Coh}}=\frac{s_{m}^\text{Coh}-\mu^\text{Coh}}{\sigma^\text{Coh}}.
\end{equation}
The overall score of $r_m$ is calculated as:
\begin{equation}
\setlength{\belowdisplayskip}{4pt} 
\setlength{\abovedisplayskip}{4pt}
    s_m = \frac{\overline{s_{m}^\text{Bert}}+\overline{s_{m}^\text{Coh}}}{2}.
\end{equation}

\paragraph{Converting to Training Labels} After obtaining $\{s_m\}_{m=1}^M$, we sort them in ascending order. We use the indices of $\{s_m\}_{m=1}^M$ in the ordered sequence as the label, denoted as $\mathbf{\pi}\in\mathcal{R}^M$. $\mathbf{\pi}$ actually represents a permutation of $\{1,2,\cdots,M\}$, indicating the priority of routing $q$ to $T_1,\cdots,T_M$, respectively. 
This kind of label enables us to create a listwise ranking loss to effectively leverage the priority relationships between each pair of tools.

\begin{table*}[!t]
\centering
\small 
\setlength{\tabcolsep}{2pt}
\begin{tabular}{@{}l|lllll|lllll@{}}
\toprule
\multirow{2}{*}{Search Strategies} & \multicolumn{5}{c|}{Qwen2-max} & \multicolumn{5}{c}{GPT4} \\ \cmidrule(l){2-11} 
 & \tiny{WebQA} & \tiny{PrivateMH} & \tiny{NLPCC-MH} & \tiny{CDQA} & \tiny{SogouQA} & \tiny{WebQA} & \tiny{PrivateMH} & \tiny{NLPCC-MH} & \tiny{CDQA} & \tiny{SogouQA} \\ \midrule
No-RAG & 4.609 & 2.77 & 1.803 & 2.543 & 3.31 & 3.927 & 1.924 & 2.647 & 2.538 & 3.715 \\ \midrule
\rowcolor{gray!10}
\multicolumn{11}{c}{Using one  search engine} \\
Quark & \textbf{4.833} & 3.38 & 2.258 & 3.691 & 4.484 & 4.321 & 2.657 & 2.735 & 3.73 & 4.42 \\
Bing & 4.79 & 3.309 & 2.151 & 3.661 & 4.437 & 4.168 & 2.461 & 2.737 & 3.559 & 4.33 \\
Google & 4.797 & 3.205 & 2.104 & 3.745 & 4.395 & 4.225 & 2.557 & 2.644 & 3.79 & 4.312 \\ \midrule
\rowcolor{gray!10}
\multicolumn{11}{c}{Using two search engines} \\
Quark+Bing & 4.823 & 3.428 & 2.274 & 3.829 & 4.546 & \textbf{4.323} & 2.669 & 2.755 & 3.877 & 4.445 \\
Quark+Google & 4.822 & 3.418 & 2.261 & 3.917 & 4.487 & 4.285 & 2.682 & 2.701 & 3.925 & 4.412 \\
Bing+Google & 4.815 & 3.363 & 2.214 & 3.858 & 4.458 & 4.24 & 2.61 & 2.765 & 3.893 & 4.36 \\ \midrule
\rowcolor{gray!10}
\multicolumn{11}{c}{Using three search engines} \\
\small{Quark+Bing+Google} & 4.808 & \textbf{3.445} & \textbf{2.352} & \textbf{3.993} & \textbf{4.584} & 4.285 & \textbf{2.705} & \textbf{2.796} & \textbf{3.962} & \textbf{4.45} \\ \bottomrule
\end{tabular}
\caption{The Correctness results under different combinations of LLM and search tools. Scores with \textbf{bold} denote the best result. Values are calculated by averaging 3 random experiments.}
\label{tab:baseline_comb}
\end{table*}

\subsection{Model Training and Inference}
Given the labeled $(q,\pi)$ examples, we train a neural model $\mathcal{M}$ to predict the scores of routing $q$ to $\{T_m\}_{m=1}^M$, respectively:
\begin{equation}
\setlength{\belowdisplayskip}{4pt} 
\setlength{\abovedisplayskip}{4pt}
    \mathbf{p} = \mathcal{M}(q)
\end{equation}
where $\mathbf{p}=\{p_1, \cdots,p_m,\cdots,p_M\}$, $p_m$ denotes the score of routing $q$ to $T_m$.
Then, we adopt the listwise ranking loss ListMLE \cite{Guiver2009Plackett} to optimize model parameters:
\begin{equation}
\setlength{\belowdisplayskip}{4pt} 
\setlength{\abovedisplayskip}{4pt}
    \mathcal{L}=\text{ListMLE}(\mathbf{p},\mathbf{\pi}).
\end{equation}
During inference, our method only requires the input $q$ to predict the scores of using each search tool, without actually calling any search tool.

\begin{figure*}[!ht]
    \centering
    \subfigure[The data scaling result when using Qwen2-max for response generation.]{
        \includegraphics[width=0.94\linewidth]{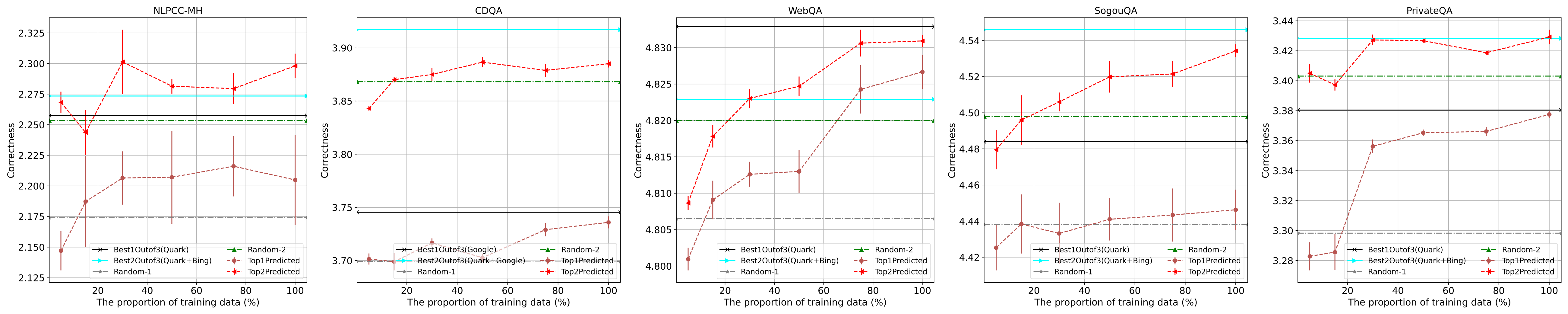}
        }
    \\   
    \subfigure[The data scaling result when using GPT4 for response generation.]{
    	\includegraphics[width=0.94\linewidth]{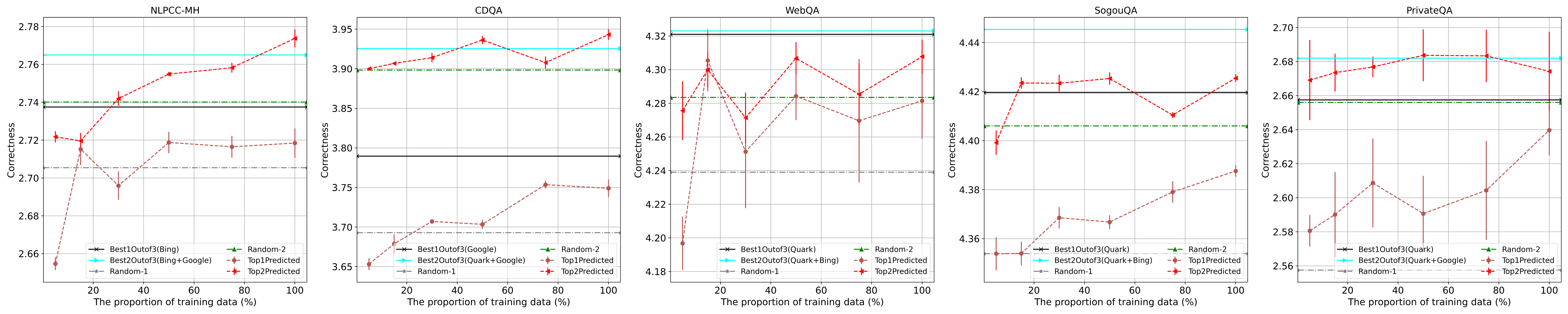}
        }
    \caption{The data scaling result of our method. For \textbf{Best1Outof3} and \textbf{Best2Outof3}, we explicitly indicate the specific search strategy in parentheses. The values are obtained by averaging 5 random experiments.}
    \label{fig:overall_scaling_effect}
\end{figure*}

\begin{figure*}[!htb]
    \centering
    \includegraphics[width=0.94\linewidth]{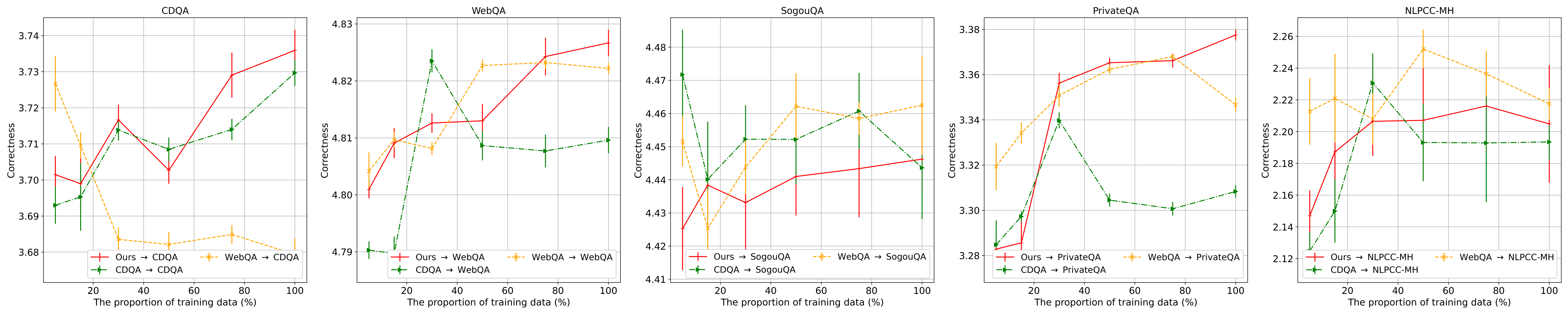}
    \caption{The evaluation result of the generalization abilities of different methods. We report the \textbf{Top1-Predicted} result. Values are calculated by averaging 5 randomized experiments. We try three different types of training examples on five test sets. "$A \rightarrow B$" means we train the model on $A$ and evaluate the model on $B$. For example, "CDQA $\rightarrow$ WebQA" means we train the model on CDQA and evaluate the model on WebQA. Therefore, "CDQA $\rightarrow$ CDQA" and "WebQA $\rightarrow$ WebQA" are actually in an i.i.d. setting.}
    \label{fig:ood_eval}
\end{figure*}

\section{Experiment}
\subsection{Experimental Settings}

\paragraph{Datasets}
We use four public open-domain QA datasets, including NLPCC-MH \cite{wang2019nlpccmh}, CDQA \cite{Xu2024LetLT}, WebQA \cite{Chen2019BidirectionalAM}, SogouQA\footnote{https://aistudio.baidu.com/datasetdetail/17514}, and a private QA dataset, named PrivateQA, for evaluation. The statistics of the dataset and the details of the construction of PrivateQA are shown in Appendix \ref{app:experimental_setting}. 

\paragraph{Search Tools and LLMs}
We consider 3 search tools, i.e., Quark, Bing, and Google, for document retrieval.
For each query, we retrieve $k=6$ documents. That is, in the upper-bound construction process, we only retain the top-ranked 2 documents for each search tool. 
We use Qwen2-max and GPT4 for response generation.

\paragraph{The Evaluation Metric}
Each test example is in the form of \textit{(query, answer)}. For each query, we use the different combinations of LLMs and search tools to obtain the retrieval-augmented responses. We adopt the \textbf{Correctness} score implemented by Llama-index \cite{Liu_LlamaIndex_2022} as the metric to evaluate the quality of generated responses. For each query, we repeatedly generate three times and calculate the average correctness score.
Since we have considered 3 search engines and 2 LLMs, we have 16 RAG strategies. 
The result is shown in Table \ref{tab:baseline_comb}. In the vast majority of strategies, using more search tools will lead to better results.

\paragraph{Data Source of Training Queries}
We utilize our in-house data to construct the training examples. After filtering out illegal samples (\cref{sec:label_metrics}), we finally obtain around 110k training examples. It should be noted that there is a considerable distribution discrepancy between the training and test examples. As a result, the evaluation of our method can be regarded as in a non-i.i.d. setting.

\paragraph{Model Training Details}
We initialize our routing model as GTE-large \cite{li2023towards}. Since our method is model-agnostic, we have also tried other backbone models, as shown in \cref{sec:ablate_study}.
We utilize the AdamW optimizer with an initial learning rate of 5e-5 and a batch size of 32. The learning rate is linearly decayed to zero with a 10\% warmup ratio over 10 training epochs. Our model is evaluated through 5 randomized experiments, and we report the average results. All experiments are conducted using the PyTorch toolkit on an Ubuntu server equipped with four V100 (32GB) GPUs.

\subsection{The Evaluation of Scalability}
Our method, which eliminates the need for manual annotation, is designed to exhibit strong scalability. This means that as we increase the volume of training data, we are expected to witness a corresponding improvement in the performance of our model. In this section, we conduct a comprehensive evaluation of the scalability of our approach.

\paragraph{Setting} We alter the proportions of training data to assess the scalability of our method. For each proportion of training data, we present the average result of five random experiments.
We consider the following baselines for comparison:
(1) \textbf{Random-1} which randomly selects a tool from Quark\footnote{https://www.quark.cn/}, Bing, or Google for RAG;
(2) \textbf{Random-2} which randomly selects a combination from Quark+Bing, Bing+Google, or Quark+Google for RAG;
(3) \textbf{Best1Outof3} which is the most effective one among Quark, Bing, or Google; 
(4) \textbf{Best2Outof3} which is the most effective one among Quark+Bing, Bing+Google, or Quark+Google.

It should be noted that \textbf{Best1Outof3} and \textbf{Best2Outof3} may correspond to different specific retrieval strategies on different datasets.
Correspondingly, for each query, our method will predict the score of each tool.
Therefore, we consider \textbf{Top1-Predicted} and \textbf{Top2-Predicted}, where \textbf{Top1-Predicted} indicates that we choose the tool with the highest predicted score, and \textbf{Top2-Predicted} indicates that we choose the two tools with the highest predicted scores. 
Ultimately, we compare \textbf{Top1-Predicted} with \textbf{Best1Outof3}, and compare \textbf{Top2-Predicted} with \textbf{Best2Outof3}.

\paragraph{Result} The result is shown in Figure \ref{fig:overall_scaling_effect}. The random baselines show an unsatisfactory performance. In addition, we have the following observations.
\begin{itemize}[leftmargin=*]
    \item As the volume of training data increased, our method has demonstrated consistent and obvious performance improvements. Since our method eliminates the necessity of manual annotation, the sustained enhancement in performance highlights the effectiveness of leveraging a larger amount of real user queries, ultimately, it is promising to obtain more refined and satisfied outcomes in diverse applications. 
    \item The effectiveness of our method is observed consistently across various LLMs and multiple datasets, underscoring the robustness and stability of our approach. The key reason is that a wide variety of real user queries are directly utilized for the learning of query routing, which can effectively guarantee data diversity.
    \item On NLPCC-MH and CDQA datasets, our method outperforms the benchmarking baseline, demonstrating that it has learned the routing mechanism to some extent. This capability enables it to assign queries to the most suitable search engine, resulting in superior performance. 
    While our method does not outperform the best baseline on some datasets, we attribute this limitation to the still insufficient volume of training data. Actually, query routing among general-purpose search engines is an extremely difficult task, since even humans struggle to determine which search engine should be selected to handle a query when they have no knowledge about black-box search engines.
    Fortunately, since we have observed the data scaling effect, we have reason to believe that if we continue to expand the data volume, our method will show even greater improvements and achieve optimal results. 
\end{itemize}

Overall, by systematically varying the size of the training dataset, we observe compelling scalability for our method, which demonstrates its potential for utilizing more data for training.

\subsection{The Evaluation of Generalization Ability}

Our model theoretically has good generalization ability.
The underlying principle is that by collecting sufficient real user queries, data diversity can be ensured, enabling the model to effectively generalize to unseen data.
Consequently, in this section, we compare our approach with existing supervised methods to assess its generalization ability.

\paragraph{Setting} 
Our method uses real user queries for training. As a comparison, we reproduce existing supervised methods \cite{shnitzer2023large,mu2024adaptive} that utilize annotated \textit{(query, answer)} data for model training. More specifically, we use the training splits of CDQA and WebQA which contain 10k and 20k examples, respectively, to construct training examples (\cref{sec:super_works}). Then, we use the obtained examples to train the routing models and evaluate them on the 5 test sets.
By altering the types of training data, we observe the scaling effect of the learned models on the 5 test sets.

\paragraph{Result} The result is shown in Figure \ref{fig:ood_eval}. 
We have the following observations.
\begin{itemize}[leftmargin=*]
    \item In the independent and identically distributed (i.i.d.) settings, specifically in the "CDQA $\rightarrow$ CDQA" and "WebQA $\rightarrow$ WebQA" scenarios, supervised methods demonstrate a pronounced scaling trend. This indicates that augmenting the volume of training data correlates with sustained performance improvements, which aligns with human intuitive understanding.
    \item Conversely, in the other non-i.i.d. scenarios of supervised methods, the addition of more training data fails to yield similar performance gains. This suggests that traditional supervised methods struggle to generalize effectively in non-i.i.d. environments. Such findings highlight a critical limitation in existing approaches. That is annotated data is difficult to obtain, which greatly limits the diversity of data and thus restricts their generalization ability.
    
    \item Our method demonstrates a pronounced data scaling trend across all datasets, although the trend on NLPCC-MH is not as obvious as the trend on other test sets.
    This is primarily due to the fact that the data scale utilized by our approach significantly surpasses that of traditional supervised methods. Such an expansive data scale effectively enhances the diversity of the training data. This characteristic underscores the innovation of our approach, which is capable of automatically curating a wide array of diverse training data, eliminating the need for both time and cost consuming human annotation. 
    
\end{itemize}

\subsection{Further Discussion}
\label{sec:ablate_study}

\begin{table}[!htb]
\centering
\small 
\setlength{\tabcolsep}{1pt}
\begin{tabular}{@{}lllll@{}}
\toprule
Metrics & CDQA & WebQA & SogouQA & PrivateQA \\ \midrule
Full (ours) & 3.736 & 4.827 & 4.446 & 3.378 \\
w/o Similarity & 3.718 & 4.806 & 4.438 & 3.286 \\
w/o Coherence & 3.698 & 4.796 & 4.422 & 3.272 \\ \bottomrule
\end{tabular}
\caption{The impact of different automatic metrics in the label construction step. Values are calculated by averaging 5 randomized trials.}
\label{tab:albation_autometrics}
\end{table}

\paragraph{The Impact of Different Evaluation Metrics}

We develop the following two ablated variants to validate the effectiveness of different metrics in the label construction step: (1) "w/ Similarity" means we only use the similarity score for label construction; (2) "w/ Coherence" means we only use the coherence score for experiments. The result is shown in Table \ref{tab:albation_autometrics}. Both "w/ Similarity" and "w/ Coherence" show a performance drop. In addition, we have the following observations.
\begin{itemize}[leftmargin=*]

    \item The performance of "w/ Similarity" surpasses that of "w/ Coherence," indicating that the similarity metric is more effective. It is important to note that the similarity metric employs multi-sourced responses as the anchor, whose reliability has been validated by our experiments. This suggests that the similarity metric is a more objective measure, resulting in labels that are more robust. In contrast, the coherence metric relies solely on the knowledge within the LLM itself. However, due to inherent knowledge limitations in LLMs, it can generate incorrect information, leading to reduced effectiveness compared to the similarity metric.
    As evidence, we observe that LLMs may generate illegal ranking outputs, such as $A>B, B>C$, but $C>A$, in the coherence score calculation. These illegal outputs account for approximately 9\% of the population. 
    \item Although not as effective as the Similarity metric, the Coherence metric contributes to our method. The best result is achieved when combining two metrics.
    One possible explanation is that Similarity and Coherence assess different aspects of the generated single-sourced responses. Specifically, similarity measures how closely a single-sourced response aligns with its upper-bound, while Coherence evaluates the rationality of a single-sourced response in relation to the input query. As a result, these two metrics can effectively complement one another.

\end{itemize}

\begin{table}[!htb]
\centering
\small 
\setlength{\tabcolsep}{1pt}
\begin{tabular}{@{}lllll@{}}
\toprule
Backbone & CDQA & WebQA & SogouQA & PrivateQA \\ \midrule
GTE-large (ours) & 3.736 & 4.827 & 4.446 & 3.378 \\
w/ Roberta-large & 3.716 & 4.826 & 4.453 & 3.343 \\
w/ Qwen-0.5B & 3.727 & 4.808 & 4.419 & 3.290 \\
\bottomrule
\end{tabular}
\caption{The result under different backbone models. Values are calculated by averaging 5 randomized trials.}
\label{tab:diff_backbone}
\end{table}

\paragraph{The Result under Different Backbone Models}
We also investigate the impact of different backbone models. Besides GTE-large, we have also evaluated Roberta-large and Qwen2-0.5B, as they possess a comparable number of parameters to GTE-large. The result is shown in  Table \ref{tab:diff_backbone}.
We observe a minor result difference between GTE-large and Roberta-large. This is because they belong to the same type of models, and after training, they will all converge towards the distribution of the training data. However, Qwen2-0.5B shows a large performance drop compared to GTE-large and Roberta-large. We speculate that such decoder-only small models are less effective for regression tasks. 
Due to time and computational resource limitations, we are unable to try larger models, such as Qwen2-4 and Qwen2-7B, for experiments.
Finally, we choose GTE-large as the backbone of our work.


\paragraph{The Quality of Generated Training Datasets}
We additionally assess the quality of the generated training datasets through manual review. We randomly sampled 200 automatically constructed examples, and manually evaluate the average accuracy of the labels generated by different metrics. The result is shown in Table \ref{tab:quality_trainset}.

\begin{table}[!htb]
\centering
\begin{tabular}{@{}ll@{}}
\toprule
Metrics    & Accuracy (\%) \\ \midrule
Similarity & 83                 \\
Coherence  & 79                 \\
Overall    & 86               \\ \bottomrule
\end{tabular}
\caption{The average accuracy of automatically labelled training datasets.}
\label{tab:quality_trainset}
\end{table}

The Similarity metric provides more accurate labels compared to the Coherence metric, as confirmed by our experiment (\cref{sec:ablate_study}). By integrating both metrics, our weighted overall metric yields the highest accuracy in labeling. Given the automated nature of our method, we consider an 86\% accuracy rate to be satisfactory.

\begin{figure}
    \centering
    \includegraphics[width=0.85\linewidth]{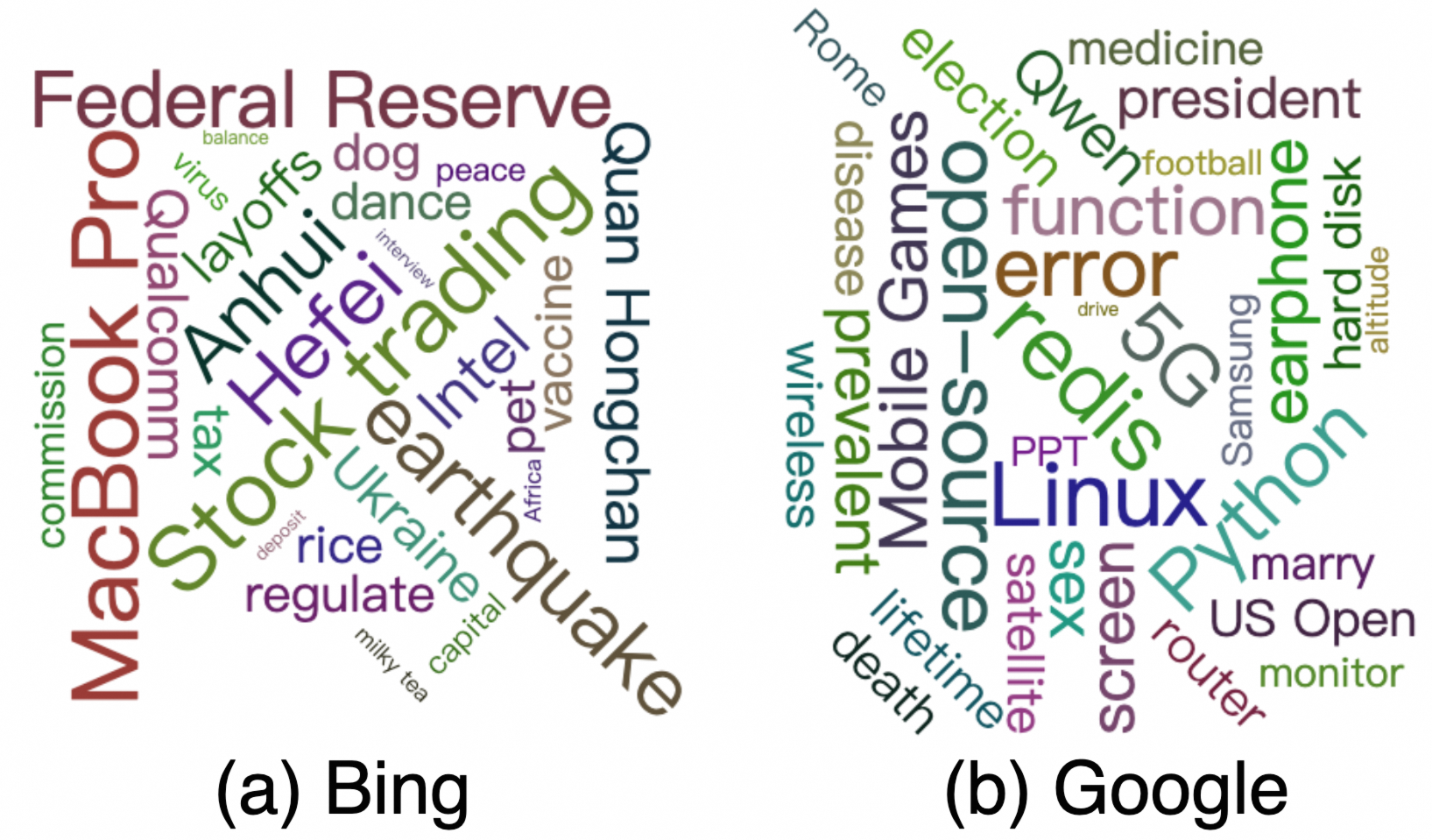}
    \caption{The visualization of the high-frequency words. The words are translated from Chinese. We leave the original version in Appendix \ref{app:zh_visual}. }
    \label{fig:visualize_wordcloud}
\end{figure}

\paragraph{Visualization}
We conduct a further analysis of the constructed training labels. According to the label $\pi$, we categorize the queries into three distinct groups: those should be assigned to Quark, Bing, and Google, respectively. Then, we tally the high-frequency words presented in the queries of each group and visualize high-frequency words in each group,
as depicted in Figure \ref{fig:visualize_wordcloud}. In the "Bing" group, terms such as \textit{"earthquake", "MacBook Pro", and "Hong Kong stock"} demonstrate a high frequency, indicating that Bing has good support for news hotspots as well as significant political and economic events.
In the "Google" group, high-frequency words include "python, open-source, 5G", etc. 
This indicates that Google is better able to handle programming, technology, and other queries that require strong professionalism. 
This is quite intuitive, as Google, renowned as the world's leading search engine, offers the most extensive internal documentation available, enabling the platform to effectively tackle a diverse array of professional issues. 

\section{Conclusion}
In this work, we present an unsupervised query routing approach. Given the insight that multi-sourced retrieval yields superior performance, we ingeniously design an automated process to assess the quality of single-sourced responses, thereby generating training labels without manual annotation. Experimental results demonstrate that our method has excellent scalability and generalization ability. We are confident that our approach will inspire new advancements within the community.

\section*{Limitations}
Due to the high costs associated with large-scale data processing, we currently have access to only approximately 110k training samples. It is important to note that query routing among general-purpose search engines presents a particularly challenging problem, and it is likely that 110k examples are still insufficient for our needs. 
In the future, given a larger budget, we plan to conduct experiments using a greater number of examples. For our methodology, we have chosen smaller models instead of larger Qwen2-7B or Qwen2-14B as the backbone for two primary reasons: (1) Query routing demands low latency, whereas larger models tend to operate at slower speeds, which does not meet our requirements. (2) Tool selection necessitates that the model accurately predicts the score for each tool in relation to each query, which effectively constitutes a numerical prediction task. However, auto-regressive LLMs often exhibit significant challenges in numerical prediction.

\bibliography{custom}
\appendix

\section{The Prompts in Our Method}
\label{app:prompts}
\begin{figure}
    \centering
    \includegraphics[width=\linewidth]{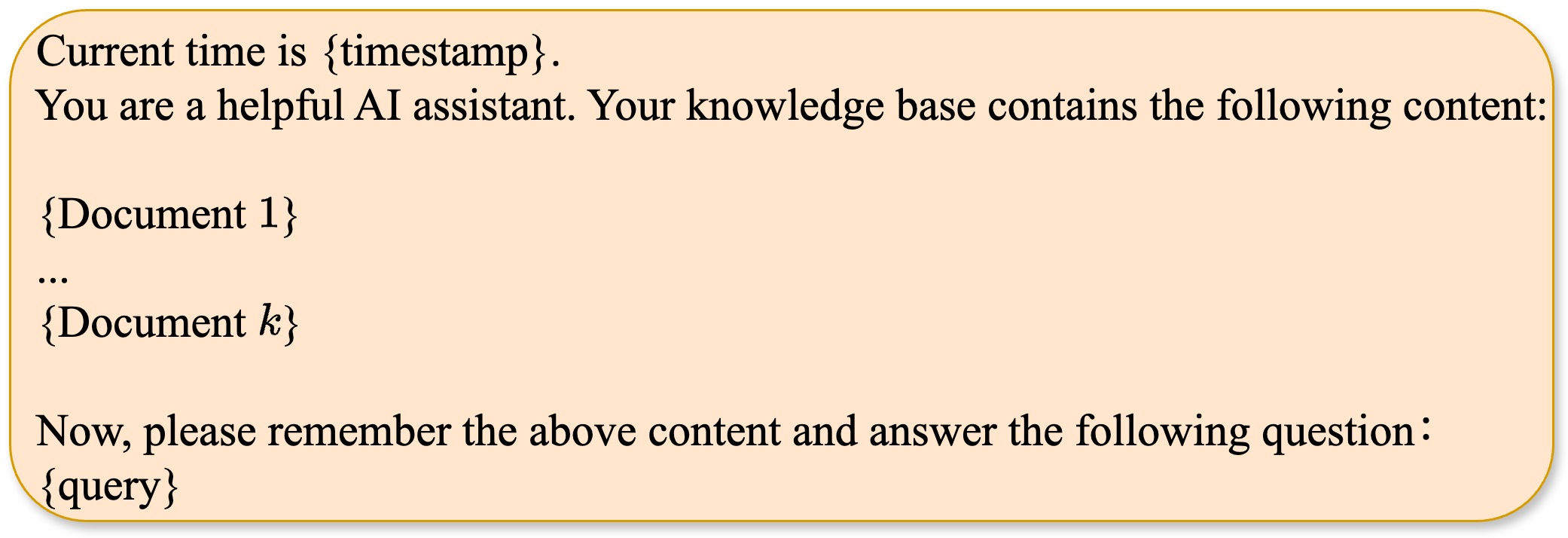}
    \caption{The prompt for response generation. The prompt is translated from Chinese.}
    \label{fig:single_rag_prompt}
\end{figure}

\begin{figure}
    \centering
    \includegraphics[width=\linewidth]{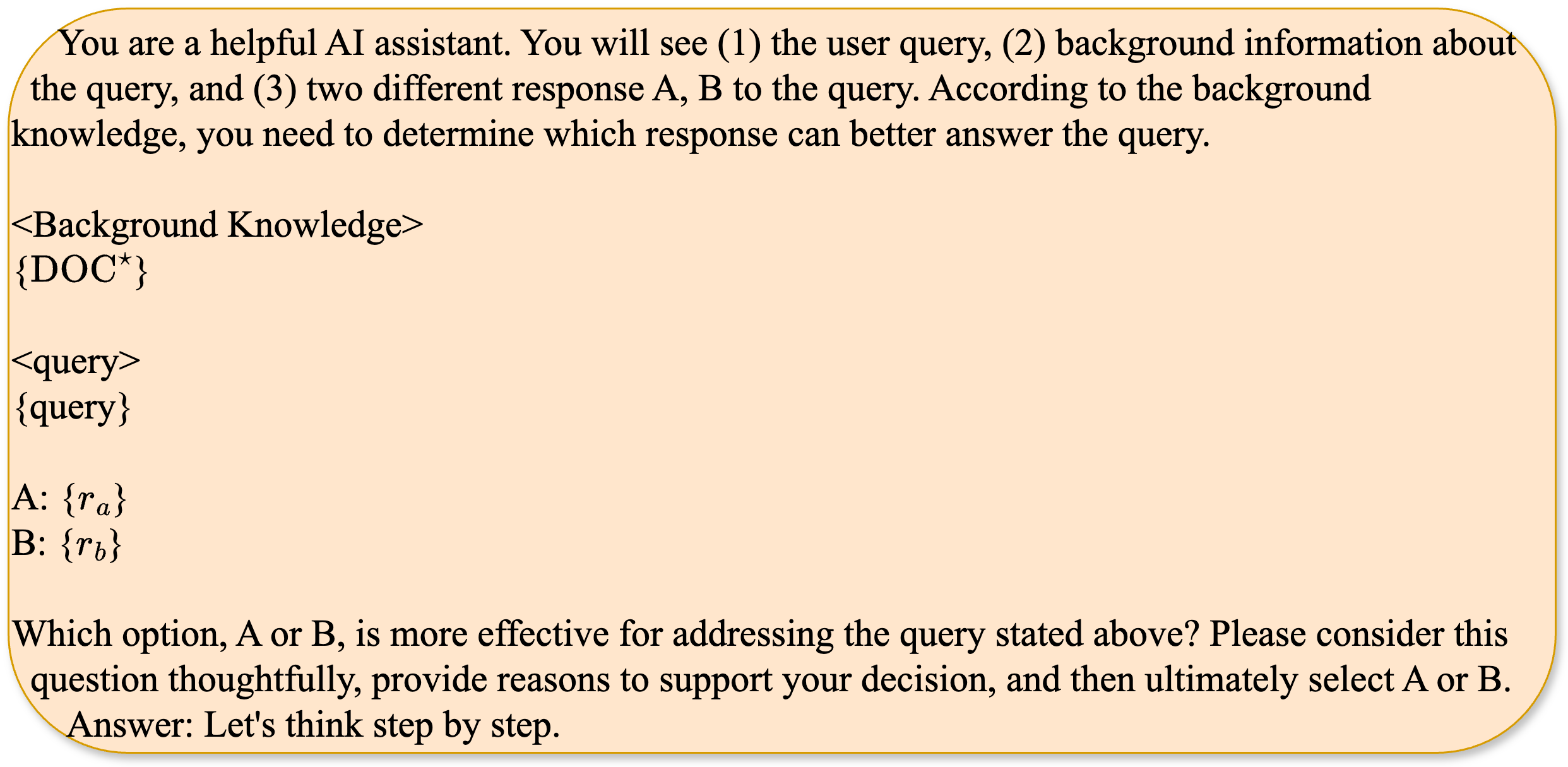}
    \caption{The prompt used for coherence evaluation. The prompt is translated from Chinese.}
    \label{fig:nli_prompt}
\end{figure}

Figure \ref{fig:single_rag_prompt} and \ref{fig:nli_prompt} show the prompts in response generation and coherence evaluation.

\section{Experimental Setting Details}
\label{app:experimental_setting}
\paragraph{Construction Details about PrivateQA}

PrivateQA primarily consists of questions related to timeliness. The question-answer examples are collected from public QA websites or translated from public English datasets using GPT4. We prioritize privacy protection and data integrity, ensuring that the dataset contains only open-domain QA content and excludes any private, harmful, or unethical material. PrivateQA is used to evaluate LLMs’ capability to answer questions about timeliness. Due to confidentiality requirements, we are unable to disclose the dataset publicly.

\begin{table}[!htb]
\centering
\small 
\setlength{\tabcolsep}{1pt}
\begin{tabular}{@{}llllll@{}}
\toprule
 & \tiny{WebQA} & \tiny{PrivateQA} & \tiny{NLPCC-MH} & \tiny{CDQA} & \tiny{SogouQA} \\ \midrule
\begin{tabular}[c]{@{}l@{}}Number \\ of samples\end{tabular} & 400 & 400 & 400 & 400 & 400 \\ \bottomrule
\end{tabular}
\caption{The datasets statistics.}
\label{tab:data_statistics}
\end{table}


\paragraph{Data Statistics}
The statistics of used datasets are shown in Table \ref{tab:data_statistics}.
The examples are randomly sampled from the original data split.

\section{The Chinese Visualization Result}
\label{app:zh_visual}
\begin{figure}
    \centering
    \includegraphics[width=0.9\linewidth]{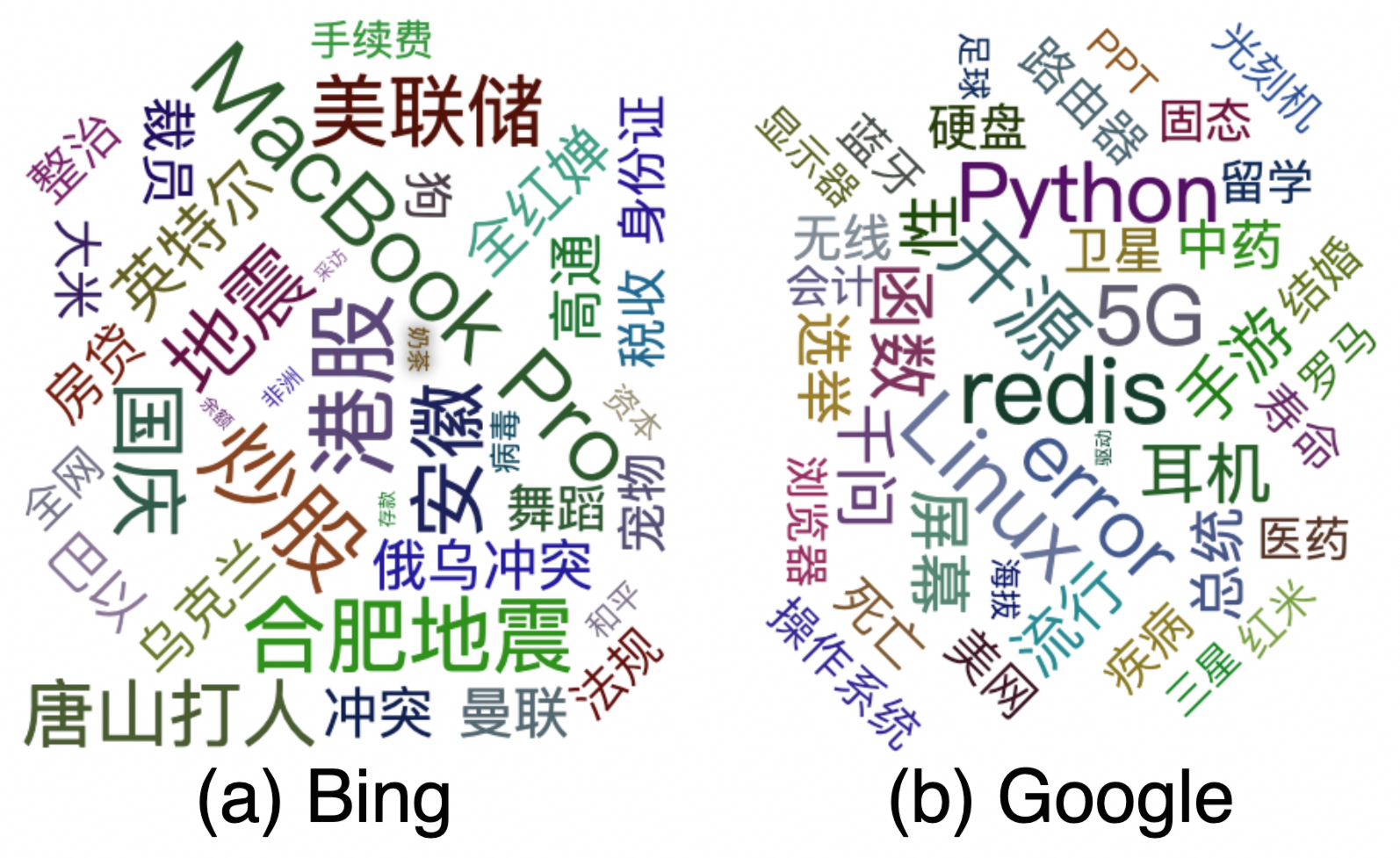}
    \caption{The Chinese visualization result of the high-frequency words.}
    \label{fig:visualize_wordcloud_zh}
\end{figure}

Figure \ref{fig:visualize_wordcloud_zh} presents the original Chinese version of the high-frequency words.

\end{document}